\documentclass[showpacs,twocolumn,aps,prb,floatfix,superscriptaddress]{revtex4}%
\usepackage{amsmath,amssymb,amsfonts,graphicx}%
 \usepackage{times}
\usepackage{amsmath}%
\usepackage{amsfonts}%
\usepackage{amssymb}%
\usepackage{graphicx}
\def\unit#1{\ \mathrm{#1}}

\begin{document}
\title{Spin-polarization of (Ga,Mn)As measured by Andreev Spectroscopy: \\The role of spin-active scattering}
\author{S. Piano}
\affiliation{School of Physics and Astronomy, University of Nottingham, University Park, Nottingham, NG7 2RD, United Kingdom}
\author{R. Grein}
\affiliation{Institut f\"{u}r Theoretische Festk\"{o}perphysik and DFG-Center for Functional Nanostructures, Karlsruhe Institute of Technology, D-76128 Karlsruhe, Germany}
\author{C. J. Mellor}
\affiliation{School of Physics and Astronomy, University of Nottingham, University Park, Nottingham, NG7 2RD, United Kingdom}
\author{ K. V\'yborn\'y}
\affiliation{Institute of Physics ASCR, v.v.i., Cukrovarnick\'a 10, Praha 6, CZ-16253, Czech Republic}
\author{R. Campion}
\affiliation{School of Physics and Astronomy, University of Nottingham, University Park, Nottingham, NG7 2RD, United Kingdom}
\author{ M. Wang}
\affiliation{School of Physics and Astronomy, University of Nottingham, University Park, Nottingham, NG7 2RD, United Kingdom}
\author{M. Eschrig}
\affiliation{Institut f\"{u}r Theoretische Festk\"{o}perphysik and DFG-Center for Functional Nanostructures, Karlsruhe Institute of Technology, D-76128 Karlsruhe, Germany}
\affiliation{Fachbereich Physik, Universit\"at Konstanz, D-78457 Konstanz, Germany}
\affiliation{SEPnet and Hubbard Theory Consortium, Department of Physics, Royal
Holloway, University of London, Egham, Surrey TW20 0EX, United Kingdom}
\author{B. L.  Gallagher}
\affiliation{School of Physics and Astronomy, University of Nottingham, University Park, Nottingham, NG7 2RD, United Kingdom}
\date{February 18, 2011}
\pacs{72.25.Mk, 74.45.+c, 75.50.Pp \vspace*{-.1cm}}

\begin{abstract}
We investigate the spin-polarization of the ferromagnetic semiconductor
(Ga,Mn)As by point contact Andreev reflection spectroscopy. The  conductance
spectra are analyzed using a recent theoretical model that accounts for momentum- and spin-dependent
scattering at the interface. This allows us to fit the data without resorting, as in the case of the  standard spin-dependent Blonder-Tinkham-Klapwijk (BTK) model, to an effective temperature or a statistical distribution of
superconducting gaps. We find a transport polarization
 $P_{C}\approx57 \%$, in considerably better agreement with the $\vec{k}%
\cdot\vec{p}$ kinetic-exchange model of (Ga,Mn)As, than the significantly larger estimates inferred from the  BTK model. The temperature
dependence of the conductance spectra is fully analyzed.
\end{abstract}
\maketitle
The quickly evolving field of spintronics, which
addresses the manipulation and exploitation of the quantum-mechanical
spin of an object, has led to an intense search for spin-polarized materials
as promising candidates for applications. Current metal spintronic
devices and most proposed semiconductor spintronic devices aim to exploit the
net spin-polarization (SP) of charge carriers to encode and/or process information. The advantage of using
ferromagnetic (FM) semiconductors is their potential to serve as
spin-polarized carrier sources and the possibility to easily integrate them
into semiconductor devices \cite{Wolf}. Recently, the family of (III,Mn)V FM
semiconductors has attracted much attention for their potential usage in
non-volatile memory, spin-based optoelectronics and quantum computation
\cite{NaturePhys}. In particular, (Ga,Mn)As, with a Curie temperature as high
as 185 K \cite{Tomas-RMP,Olejnik} is one of the most prominent materials  for
such applications.

The degree of SP is the key parameter for most spintronic
functionalities. The injection of spin-polarized currents from FM
semiconductors into nonmagnetic semiconductor devices has been demonstrated by
measurements of the optical polarization of light emitted after recombination
of spin-polarized holes with electrons in nonmagnetic semiconductors
\cite{Ohno}. However, the resulting polarization is strongly dependent on the
experimental set-up and the spin relaxation rate in the non-magnetic part of
the device, so that it is very difficult to infer reliable values of
SP from such measurements \cite{Adelmann}. Specialized
techniques based on superconductor (SC)/FM junctions have been employed
frequently in recent years to obtain information on the SP in
metals \cite{Merservey, Woods}. In particular, the point contact Andreev
reflection (PCAR) technique has become a popular tool to measure the transport
SP $P_{C}$ of carriers at the Fermi level in FM materials
\cite{referee,Soulen}.
\begin{figure*}[tb]
\centering  \includegraphics [width=18cm]{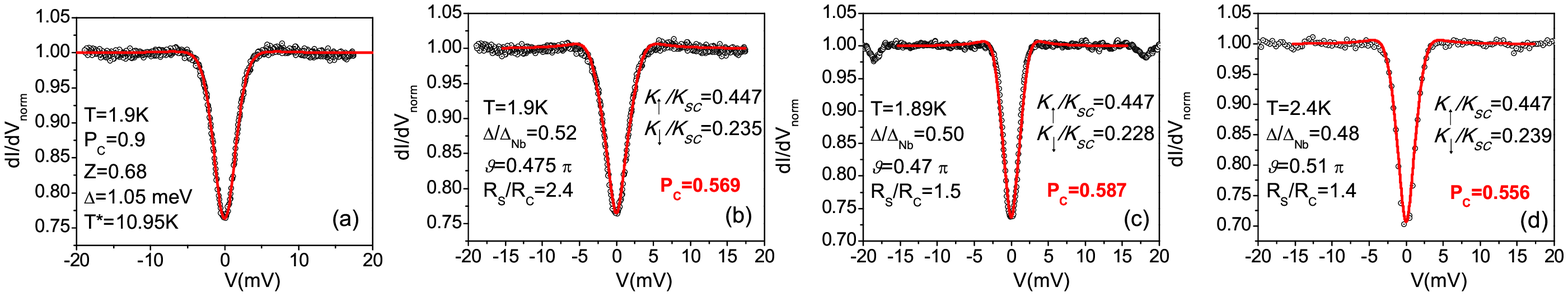}
\caption{(Color online) Experimental PCAR data (black dots) for the
  as-grown sample together with the theoretical
fit (red line), (a) using the modified BTK model, (b) according to the
spin-active scattering (SAS) model in Ref. \cite{Grein}. (c) Annealed
sample and (d) another PCAR probe applied to the as-grown sample, both
analyzed using the SAS model. The fitting parameters are in the caption of the Figs. We emphasize that the temperature (T) and
the value of the transport SP $P_{C}$ in Figs. (b), (c) and (d) are not fitting parameters, as explained in the text.}%
\label{LowTemp}%
\end{figure*}
The PCAR data contains information regarding the carrier SP
because the Andreev reflection process is spin-sensitive. Andreev reflection
is the only allowed process of charge transfer across a contact between a
SC and a normal metal at energies below the SC gap. In this
process, a Cooper pair is created in the SC when a quasiparticle tunnels from
the normal metal and a hole is reflected back coherently. Two
particles with opposite spin are in fact transferred, hence the probability of
this process  is reduced when the density of states for spin-$\uparrow$/$\downarrow$
carriers is different. This results in a suppression of the conductance for
voltages below the SC gap and thus the degree of
SP can be estimated from such conductance spectra
\cite{Strijkers, Mazin}.
However, the reliability of the $P_{C}$-values obtained with
this technique ---  based on fitting the spectra to an extension of the
Blonder-Tinkham-Klapwijk (BTK) model \cite{BTK, Mazin} --- is currently under
debate \cite{Woods, Xia, Grein,Geresdi}.

Here we present a detailed PCAR study of the carrier (hole) SP in
(Ga,Mn)As. The measurements have been carried out at different temperatures
starting from about 2K up to the point where the SC gap in the tip closes. The
shape of the experimental conductance spectra shows a suppression of the
conductance for voltages smaller than the SC gap and a complete absence of
coherence peaks at the gap edge. However, in agreement with earlier PCAR
measurements on this material, the conductance suppression below the gap
energy is rather small, with a minimum of about $0.7\ G_{N}$, where $G_{N}$ is
the conductance in the normal state. This feature would usually hint at an
intermediate SP of the FM-material. Yet, if one tries to fit
these spectra with the BTK theory, it turns out that the two key features, the
absence of coherence peaks and the rather high zero-bias conductance, cannot
be reconciled within this model. To remedy this situation, one needs to appeal
to an ``effective'' fictitious temperature, $T^{\ast}$, which we find to be almost 6
times as high as the real sample temperature in our case. With this
additional fit parameter,  satisfactory fits can be achieved extracting a high value of $P_C$ ($\gtrsim 90\%$).

We propose a more sensible interpretation of the data using a model of
interface scattering that goes beyond the BTK theory, showing that an excellent
agreement can be achieved \emph{without} an effective temperature. From this
analysis we infer a value of the SP of about $57\%$. The expected range of $P_C$ is moreover analyzed with the $\vec {k}\cdot\vec{p}$ exchange model, providing further confirmation that the BTK SP value was far from realistic. We also
notice a substantial reduction of the SC energy gap and critical
temperature of the Nb tip,  probably related to an inverse magnetic
proximity effect. From measurements of the conductance spectra at different
temperatures, we extrapolate the temperature dependence of the energy gap.

We have investigated 7\% Mn
doped, $25$nm thick (Ga,Mn)As samples. They are grown on a $400$nm
thick, highly carbon-doped ($\approx 10^{19}$ cm$^{-3}$) buffer layer
to minimise series resistance. The Curie temperature $T_C$ is $\approx70$~K
for the as-grown sample and $\approx140$~K after 24~h of annealing at 190$\unit{^\circ C}$.
The details of the
sample growth and preparation are described elsewhere
\cite{preparation}. The experiments were carried out by means of a
variable temperature (1.5--300~K) cryostat. Sample and Nb tip
(chemically etched) were introduced into the PCAR probe, in which a
piezo motor and scan tube can vary the distance between tip and
sample. The PCAR junctions were formed by pushing the Nb tip on the
(Ga,Mn)As surface with the probe thermalized in $^{4}$He gas. The
current-voltage $I$ vs $V$ characteristics were measured by using a
conventional four-probe method and, by using a small ac modulation of
the current, a lock-in technique was used to measure the differential
conductance $\mathrm{d}I/\mathrm{d}V$ vs $V$.

In Fig.~\ref{LowTemp} we show conductance spectra at low temperature
($T \approx2$ K), the resistance at high bias (that corresponds to the
resistance of the normal state) for the different contact was about
$28-30$ $\Omega$.  The data of Fig.~\ref{LowTemp}(a) and (b) is
identical, but (a) is fitted with the extended BTK model and b-d with
the theory of \cite{Grein}. The data has been normalized using the
background conductance estimated at large voltage ($V
\gg\Delta_{\mathrm{Nb}}/e$) regions, where $\Delta_{\mathrm{Nb}}
\approx 1.5$~meV is the SC gap of Nb. All conductance spectra show a
moderate dip and completely suppressed coherence peaks at the gap
edge. While different PCAR probes on the same sample,
Fig.~\ref{LowTemp}(b,d), result in different point-contact resistances
$R_{pc}$, the fitted value of $P_C$ is almost constant. No significant
difference in the spectra (and $P_C$) has been noticed before and
after annealing as shown in Fig.~\ref{LowTemp}(b--d), which is
remarkable given that such annealing reduces the resistivity by a
factor of $ \sim 2$ in these samples due to the large increase in hole
density.

To fit the experimental data in Fig.~\ref{LowTemp}(a), we have used as free
parameters: $P_{C}$; the strength of the barrier, $Z$; the SC
energy gap, $\Delta$; and $T^{\ast}$ and infer
the SP of about $90\%$, consistent with the other
values reported in literature and a reduction of the SC energy
gap. We underline that using the BTK model requires a very high effective
temperature, $T^{\ast}=10.95$ K, which is more than $5$ times  the
measured temperature of $1.9$ K. According to
[\onlinecite{Panguluri}], this effective temperature accounts for
inelastic scattering in the (Ga,Mn)As sample, but in any case it is a
parameter introduced ''ad hoc'', and whether such a high value of $T^{\ast}$
can be justified on this basis is not clear \cite{gamma}.

Recently, a theoretical model was
introduced which allows for a more realistic description of interface
scattering in the calculation of charge and spin transport across such
point contacts \cite{Grein, Eschrig2009}. When a contact with a
magnetic material is created, one would expect that the scattering
properties of quasiparticles depend on their spin. When no tunneling
potential is present, the transparency of the interface is controlled
by wave vector mismatches. Since wave vectors of $\uparrow$- and
$\downarrow$-spin quasiparticles are different in the FM material,
their transmission probabilities should differ accordingly.  Moreover,
it was shown that scattering phases can play an important role in this
case \cite{Eschrig2008}. While a global phase will never affect any
physical properties, the relative phase difference, that
quasiparticles incident from the SC may acquire upon being reflected
at the magnetic interface, induces a triplet proximity effect and
leads to substantial modifications of conductance spectra
\cite{Grein}. This relative scattering phase is called
\emph{spin-mixing angle} or spin-dependent interface phase shift. In
the case of point-contacts, it suppresses the coherence peaks at the
gap-edge and shifts their spectral weight to energies below the gap,
where interface Andreev bound states are induced. This mechanism
allows for an alternative interpretation of the PCAR spectra analyzed
here. Using a minimal model of spin-active scattering (SAS), we show
that excellent fits can be achieved without resorting to the effective
$T^{\ast}$.

\begin{figure}[ptb]
\centering  \includegraphics [width=0.85\columnwidth]{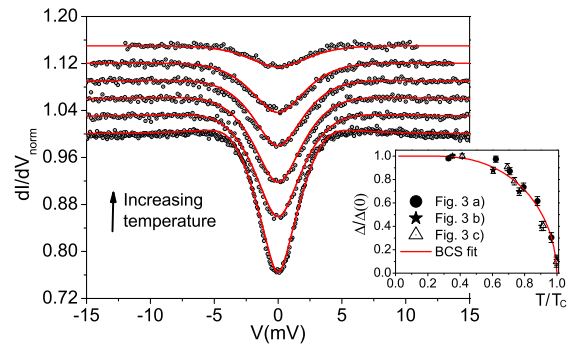}
\caption{(Color online) Spectra as a function of temperature (black dots)
with the theoretical fittings obtained by considering the spin-mixing effect
with the energy gap as only free parameter. Spectra are shifted and shown for
temperatures between $1.9$~K and $5.6$~K. Inset: Temperature dependence of the
superconducting energy gap as found from the theoretical fittings rescaled
to the BCS relation.}%
\label{Temperature}%
\end{figure}

The transport theory proposed in Ref.~\cite{Grein} relies on the normal-state
scattering matrix of the interface as a phenomenological parameter. This
S-matrix generally depends on the impact angle of the incident quasiparticles.
We assume that spin-flip scattering due to spin-orbit coupling can be
neglected (this approximation is appropriate for sufficiently strong SP), hence the S-matrix is diagonal in spin-space, yet we allow for
different transmission probabilities $t_{\uparrow}\neq t_{\downarrow}$ and a
spin-mixing angle $\vartheta$. Since there is no insulating interlayer, we
assume that the transmission probability is controlled by wave vector
mismatches. We use the averaged Fermi wave vectors of the (Ga,Mn)As spin bands
$k_{\uparrow,\downarrow}/k_{SC}$ as fit parameters. Here, $k_{SC}$ is the
Fermi wave vector in the SC. $t_{\uparrow}(\theta)$ and $t_{\downarrow}%
(\theta)$ are then calculated for any impact angle $\theta$ by wave function
matching. The density of states $N_{\uparrow,\downarrow}$ (at Fermi energy
$E_{F}$) of the respective spin-band is assumed to be proportional to
$k_{\uparrow,\downarrow}$ and independent of energy on the relevant scale of
the SC gap. The third parameter describing the interface is the spin-mixing
angle $\vartheta$ which also depends on the impact angle. If the conduction
bands of the materials are assumed to be unperturbed at the interface, this
relative scattering phase will not occur. However, if the transition from one
material to the other is smoothed on the scale of the Fermi-wavelength in Nb
\cite{Grein}, it will. We assume that this mixing phase is maximal for
perpendicular impact and goes to zero for grazing trajectories. For
definiteness, this is modeled by $\vartheta(\theta)=\vartheta\cdot\cos
(\theta)$, but even if $\vartheta(\theta)=const$ is assumed, it does not
change anything about our conclusions, since grazing trajectories contribute
little to the total conductance. Additionally, the value of the SC gap and a
spread resistance are fitted, while the temperature is taken from experiment.
For details of the calculations involved we refer the reader to Ref.
\cite{Grein, Eschrig2009}.
In addition to $\vartheta$, the fits in Fig.~\ref{LowTemp}(b-d) using
the SAS model involve $\Delta$, $k_\uparrow/k_{SC}$,
$k_\downarrow/k_{SC}$ and $R_s/R_{pc}$ as fit parameters. Here, the
spread resistance $R_{s}$ arises from the resistance of the sample
between the junction and one of the measuring contacts~\cite{Woods}
renormalizing both the voltage that drops across the contact and the
normalized conductance.  The value of $R_{s}/R_{pc}$ found by fitting
(between $2.4$ and $1.4$ in Fig.~\ref{LowTemp} (b)--(d)) is rather
high as the conductance of (Ga,Mn)As is low compared to metallic
samples and it is likely that multiple shunted contacts are
established when the tip is pressed into the sample. Spin-mixing
angles are close to $0.5\pi$ in all cases. We also find a reduction of
$\Delta$, namely to $\approx 50\%$ with respect to the zero temperature
bulk value, reported to be $1.5$~meV in Nb for the lowest temperature
spectra we measured. Judging by the disappearance of all SC features,
$T_{c}$ is reduced to $5.4-5.8$~K. This implies a deviation from the
theoretical strong-coupling BCS ratio \cite{BCS} for Nb
($2\Delta/k_{B}T_{C}\approx3.95$), instead we find
$2\Delta/k_{B}T_{C}\approx(3.3\pm0.3)$. The fitting for different
temperatures at the same measurement location was done by only varying
$\Delta$, all other parameters are kept constant. Remarkably, the
quality of the fits for all temperatures in Fig.~\ref{Temperature}
is still very good and rescaling the obtained gap values to the BCS
relation also yields a satisfactory agreement, see inset of
Fig.~\ref{Temperature}. The strong reduction of the gap could indicate that an inverse
proximity effect may be important.

\begin{figure}[ptb]
\includegraphics[width=0.9\columnwidth]{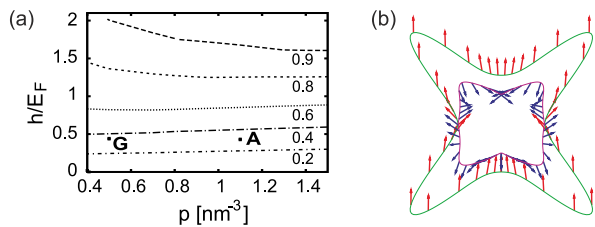} \caption{
  (Color online) The
  $\vec{k}\cdot\vec{p}$ model of (Ga,Mn)As~\cite{Abolfath} used to evaluate (a)
  total $P_{C}$ calculated by generalizing Eq.~(\ref{eq-01}) to all
  six bands involved (only isolines of $P_C$ are shown) and (b)
  expectation values of spin (spin-textures) in the heavy-hole bands
  at the Fermi surface ($k_{y}$, $k_{z}$ section is shown;
  $h=165\unit{meV}$, $p=0.8$~nm$^{-3}$). Note that, while the
  $z$-component of spin is not a good quantum number, the two bands
  can still be approximately labeled as "up" (majority) and "down"
  (minority).}
\label{fig-03}%
\end{figure}

Wavevectors inferred from the fits in
Fig.~\ref{LowTemp}(b)--(d) vary little around
$k_{\uparrow}/k_{SC}=0.447$ and $k_{\downarrow}/k_{SC}=0.237$.
Based on the assumptions of our model \cite{Grein, Eschrig2009},
the transport SP
\begin{equation}
P_{C}=2\,[\langle s_{z}v\rangle_{\uparrow}+
              \langle s_{z}v\rangle_{\downarrow}]\,/\,[\langle
v\rangle_{\uparrow}+\langle v\rangle_{\downarrow}]\,,\label{eq-01}%
\end{equation}
can be expressed as $P_{C}\approx(1-r^{2})/(1+r^{2})$ with
$r=k_{\downarrow}/k_{\uparrow}$, provided that  the effective masses in both
spin-bands $\{\uparrow,\downarrow\}$ are approximately the same.
In the Fermi-surface averages $\langle f\rangle_{i}\equiv\int
d^{3}kf(\vec{k})\delta(E_{F}-\varepsilon_{\vec{k},i})$, we consider
$v$ and $s_z$ that are the group velocities $(1/\hbar)|\nabla_{k}\varepsilon
_{\vec{k},i}|$ and the spin expectation value projected to the direction of
magnetization (taken to be $z$ here). $\varepsilon_{\vec{k},i}$ are
the band dispersions and $E_F$ is the Fermi level. In the following,
we comment on two particular aspects of the values of $P_C$ deduced
from the PCAR spectra: (i) our annealed and as-grown samples lead us,
within experimental error, to the same value of $P_C$ which (ii)
is significantly lower than previously assumed based on
the BTK theory.
We explain these findings in the framework of the warped six-band
$\vec {k}\cdot\vec{p}$ model with mean field kinetic-exchange due to
Mn~\cite{Abolfath}. It allows us to evaluate an estimate for $P_C$ in Eq.~(\ref{eq-01})
(extended to six bands) using essentially two parameters: $E_F$ and
FM splitting $h$. In Fig.~\ref{fig-03}(a) we recast the
former into the total carrier (hole) concentration $p$ for convenience
while the latter is a product of $N_{\mathrm{Mn}}$, concentration of
Mn moments that participate in the FM order, Mn total spin
$S_{\mathrm{Mn}}=\frac{5}{2}$ and the Mn-hole exchange $J_{pd}$.
Points labelled 'G' (as-grown) and 'A' (annealed) correspond to $h$ determined from the
commonly accepted value \cite{Okabayashi:1998_a} of $J_{pd}=55\unit{meV\cdot nm^3}$ and the
remanent magnetization at low temperatures ($T\ll T_C$) as measured by
SQUID magnetometry for the as-grown and annealed sample,
respectively \cite{note}.  Addressing the points (i) and (ii) above, we note
that $P_C$ has nearly equal values in 'G' and 'A' and these values are
rather low. In particular, (i) the increase of hole
concentration upon annealing (as also witnessed by the drop of $\rho$)
is compensated for by an increase of the FM splitting which
may result in practically unchanged $P_C$ after annealing; the values $P_C \approx 0.38$
and $0.31$ ('G' and 'A' in Fig.~\ref{fig-03}(a)) should be viewed as
equal within the experimental uncertainty of $\Delta
p=0.2\unit{nm}^{-3}$ (hole concentration is not precisely known as other compensation mechanisms than
interstitial Mn \cite{note} may be at works).
 Next, (ii) although we do not attempt an accurate quantitative comparison between the
theoretically estimated $P_C$ and the PCAR-inferred values,
most importantly we find that it is impossible to obtain values of
polarization approaching $90\%$ from the $\vec {k}\cdot\vec{p}$ + exchange model for reasonable hole and moment densities for our material. The remaining discrepancy
between  the $\vec {k}\cdot\vec{p}$ ($P_C\approx 35\%$) and the SAS
models ($P_C\approx 57\pm 2\%$, Fig.~\ref{LowTemp}(b-d))
may have its origin on either side. The value of $J_{pd}$
could be in fact somehow larger thus shifting the points 'A' and 'G'
upwards in Fig.~\ref{fig-03}(a). Also, the precise role of impurity
scattering in the strongly doped materials has not been elucidated so far.~\cite{Masek:2010_a}
Alternatively, the SAS model
could be refined to account for non-spherical bands that entail appreciable spin-orbit interaction. As exemplified in
Fig.~\ref{fig-03}(b), both these effects are significant in
(Ga,Mn)As. Nevertheless, even with these imperfections, the combined
results of the SAS analysis of the PCAR data and the $\vec
{k}\cdot\vec{p}$ model provide convincing evidence that the transport SP in typical
high-$T_C$ (Ga,Mn)As sample is significantly less than $~100\%$.

We have studied the transport spin-polarization at the
Fermi level in (Ga,Mn)As with the PCAR technique, using a recently developed
theory that accounts for spin-active scattering at the interface to model the
experimental results. Compared to previous work on PCAR with this material,
this allowed us to drop the assumption of an effective temperature. The value
of the SP we obtain from this analysis is sizeable but
significantly smaller than that inferred by earlier studies and it now agrees
better with predictions of the $\vec{k}\cdot\vec{p}$ kinetic-exchange model of
(Ga,Mn)As. We also investigated the full temperature dependence of the spectra
and find a strong suppression of the SC gap. The temperature dependence of the
fitted gap values is in agreement with the BCS relation.
Our study paves the way for further investigations to improve the understanding of transport mechanisms that occur at SC/FM-semiconductor interfaces, whose satisfactory characterization stands as a topic of central interest both for fundamental physics and for technological applications in spintronics. The results of this paper are likely to stimulate a critical reassessment of previous literature where the spin-active mechanism was incompletely modeled leading to unreliable estimates of SP of such materials.

We thank L. Cohen, T. Jungwirth, K. Yates, M. Humphry, and L. Kripner for discussions and support. Funding from the EU (Grants  NAMASTE-214499 and  SemiSpinNano-237375) and from the ASCR (GAAV Contract KJB100100802
and Pr\ae mium Academi\ae) is acknowledged.

\end{document}